\documentclass[aps,showkeys,twocolumn,showpacs,preprintnumbers,amsmath,amssymb,superscriptaddress,floatfix,nofootinbib]{revtex4}
\usepackage{graphicx,color,dcolumn,booktabs,bm}
\usepackage{longtable,lscape}
\usepackage{txfonts}
\usepackage{overpic}
\usepackage{epsfig}
\usepackage{amssymb}
\usepackage{rotating}
\usepackage{epstopdf}
\usepackage{indentfirst}
\usepackage{feynmf}   %{feynmp}
\usepackage{slashed}  %for Feynman symbols
\usepackage{cases}
\usepackage{color}
\usepackage{multirow}
\usepackage{graphicx,color,dcolumn,booktabs,bm}
\usepackage{cases}
\usepackage{array}

\begin{document}

\title{Masses of open charm and bottom tetraquark states in a relativized quark model}

\author{Qi-Fang L\"{u}} \email{lvqifang@ihep.ac.cn}
\affiliation{Institute of High Energy Physics, Chinese Academy of Sciences, Beijing 100049, China}
\author{Yu-Bing Dong} \email{dongyb@ihep.ac.cn}
\affiliation{Institute of High Energy Physics, Chinese Academy of Sciences, Beijing 100049, China}
\affiliation{Theoretical Physics Center for Science Facilities (TPCSF), CAS, Beijing 100049, China}

\begin{abstract}

We study the masses of open charm and bottom tetraquark states within the diquark-antidiquark scenario in the relativized quark model proposed by Godfrey and Isgur. The diquark and antidiquark masses are firstly solved by relativized quark potential, and then treated as the usual antiquark and quark, respectively. The masses of tetraquark states are obtained by solving the Schr\"{o}dinger-type equation between the new diquark and antidiquark. We find the masses of $sq\bar b\bar q$ tetraquark configuration are much higher than that of $X(5568)$. This conclusion disfavors the possibility of $X(5568)$ as a tetraquark state within the diquark-antidiquark scenario.  Further experimental searches are needed to clarify the nature of the signal observed by D0 collaboration.

\end{abstract}
\maketitle

\section{Introduction}{\label{introduction}}

Recently, the D0 Collaboration reported the evidence of a narrow structure $X(5568)$ in the $X(5568) \to B^0_s \pi^\pm$ decay process based on the $pp$ collision data at $\sqrt s = 1.96~\rm{TeV}$~\cite{D0:2016mwd}. Given the final states, the $X(5568)^+$ should have four different quark flavors $su\bar b \bar d$. Its mass and width are $5567.8\pm2.9^{+0.9}_{-1.9}~\rm{MeV}$ and $21.9\pm6.4^{+5.0}_{-2.5}~\rm{MeV}$, respectively. Assuming the final $B_s^0 \pi^+$ in $S$ wave, the quantum number is $I(J^P)=1(0^+)$. Other possibility is that the structure decays through the chains of $B_s^{0*} \pi^+$, $B^0_s \gamma$, where the soft photon is not detected. In the later situation, the quantum number of the new state would be $I(J^P)=1(1^+)$, and the mass is shifted by adding the mass difference $m(B^*_s)- m(B_s) = 48.6^{+1.8}_{-1.6} ~\rm{MeV}$. Whereafter, the LHCb Collaboration analysed the $B_s^0 \pi^+$ invariant mass distribution of $pp$ collision data at $\sqrt s = 7$ and 8 TeV, however, they found no significant excess corresponding to the claimed $X(5568)$ state~\cite{Aaij:2016iev}. Also, the CMS Collaboration failed to observe the $X(5568)$ structure in the $B_s^0 \pi^+$ invariant mass distribution~\cite{CMS:2016fvl}.

The experimental efforts have immediately attracted great interests and many theoretical studies on the $X(5568)$ with different interpretations. Considering the mass, production, strong decay, and decay constant, many investigations regard the $X(5568)$ as a tetraquark state within the framework of QCD sum rule and simple quark model~\cite{Lebed:2016yvr,Agaev:2016mjb,Wang:2016tsi,Wang:2016mee,Zanetti:2016wjn,Chen:2016mqt,Agaev:2016ijz,Liu:2016ogz,Dias:2016dme,Wang:2016wkj,Stancu:2016sfd,Tang:2016pcf,Ali:2016gdg,Agaev:2016srl,Goerke:2016hxf,Agaev:2016ifn,Agamaliev:2016wtt}, which explantation is suggested by D0 Collaboration. The structures under $B_s \pi$, $B\bar K$ and $B^* \bar K$ molecular pictures, dynamically generated states, and hybridized tetraquarks are also proposed ~\cite{Xiao:2016mho,Agaev:2016urs,Albaladejo:2016eps,Chen:2016npt,Kang:2016zmv,Lang:2016jpk,Chen:2016ypj,Lu:2016kxm,Sun:2016tmz,Esposito:2016itg}. Moreover, some works discuss the charmed partners of $X(5568)$~\cite{Agaev:2016lkl,He:2016yhd,He:2016xvd}. In addition, the non-resonance interpretations of the $X(5568)$ structure also exist. In Refs.~\cite{Liu:2016xly,Yang:2016sws}, the authors suggest that $X(5568)$ may be resulted from the near threshold kinematic effects. The quite large production rate of $X(5568)$ cannot be understood by the general hadronization mechanism~\cite{Jin:2016zuy}. It should be mentioned that the comprehensive discussions and reviews on  $X(5568)$ are performed in Refs.~\cite{Burns:2016gvy,Guo:2016nhb,Chen:2016spr}.

Before the observation of $X(5568)$, there have been some studies on open charm and bottom tetraquark states, which mainly focus on their masse spectra
~\cite{Bracco:2005kt,Maiani:2004vq,Vijande:2006hj,Carlucci:2007um,Zhang:2006hv,Gerasyuta:2008ps,Ebert:2010af,Kolomeitsev:2003ac,Zhang:2006ix,Liu:2009uz,Zhang:2009pn,Feng:2011zzb}. Those calculations include diquark-antidiquark picture, compact tetraquark, mixture of quark-antiquark and four quark components, and molecular sceneario, which intends to reveal the nature of $D_{s0}^*(2317)$. Some calculations indicate the $D_{s0}^*(2317)$ can be treated as a tetraquark state~\cite{Bracco:2005kt,Maiani:2004vq}, while others give much higher masses of the four quark components than that of $D_{s0}^*(2317)$ as well as the $DK$ threshold~\cite{Vijande:2006hj,Carlucci:2007um,Zhang:2006hv,Gerasyuta:2008ps,Ebert:2010af}. For the molecular scenario, most works indicate that a weekly bounded $DK$ state can be obtained~\cite{Kolomeitsev:2003ac,Zhang:2006ix,Liu:2009uz}, while others suggest the attraction between these two pseudoscalar mesons is not strong enough to form bound state~\cite{Zhang:2009pn}. In the open bottom sector, higher masses are given in diquark-antidiquark pictures~\cite{Ebert:2010af}, and it is found that the $B\bar K$ system can be weekly bounded~\cite{Zhang:2006ix,Feng:2011zzb}.

Unlike the $D_{s0}^*(2317)$, the $X(5568)$ cannot be regarded as a conventional meson or the mixture of quark-antiquark and four quark components due to its four distinct quark flavors. Most of the molecular and dynamically generated states interpretations cannot give the right mass and therefore can be excluded~\cite{Kolomeitsev:2003ac,Zhang:2006ix,Feng:2011zzb,Agaev:2016urs,Albaladejo:2016eps,Chen:2016npt,Kang:2016zmv,Lang:2016jpk,Chen:2016ypj,Lu:2016kxm,Burns:2016gvy}. However, the authors argue that the $B\bar K$ and $B_s \pi$ interactions can generate the $X(5568)$ dynamically when the next-leading order Lagrangian is included~\cite{Sun:2016tmz}. The tetraquark explanation is supported by QCD sum rule and simple quark model~\cite{Agaev:2016mjb,Wang:2016tsi,Wang:2016mee,Zanetti:2016wjn,Chen:2016mqt,Agaev:2016ijz,Liu:2016ogz,Dias:2016dme,Wang:2016wkj,Stancu:2016sfd,Tang:2016pcf}, but disfavored by the relativistic calculation and general discussions~\cite{Burns:2016gvy,Guo:2016nhb,Ebert:2010af}. Hence, it is natural to study the open charm and bottom tetraquark masses within a more realistic potential model, which is helpful to disentangle this conflict.

In this work, we apply the relativized quark model to calculate the masses of diquark and tetraquark states. The relativized quark model,  proposed by Godfrey and Isgur, has been extensively used to predict the properties of the conventional mesons~\cite{Godfrey:1985xj,Godfrey:1998pd,Godfrey:2015dia,Ferretti:2013faa,Ferretti:2013vua,Lu:2014zua,Ferretti:2015rsa,Godfrey:2015dva,Song:2015nia}. It has been concluded that this model gives a unified description of the light mesons, heavy-light mesons and heavy quarkonium, and therefore, it is suitable to deal with the $X(5568)$ state, in which
both light-light and heavy-light systems are included. Moreover, the relativistic effects are also considered in the model,
which may be essential for the light quarks.  We perform a calculation in the diquark-antidiquark picture following the route proposed by Ebert, Faustov, and Galkin~\cite{Ebert:2010af,Ebert:2005nc,Ebert:2007rn,Ebert:2008kb,Ebert:2008id,Monemzadeh:2014kra,Hadizadeh:2015cvx,Lu:2016cwr}. The corresponding diquark and antidiquark masses are estimated with the relativized potential firstly, an then treated as the usual antiquark and quark, respectively. The masses of the tetraquark states are, therefore,  obtained by solving the Schr\"{o}dinger-type equation between the diquark and antidiquark. In Ref.~\cite{Capstick:1986bm}, Capstick and Isgur adopted the same relativized potential to evaluate the baryon spectra within three body calculations. The computational schedule is very complicated, and can be hardly extended to study the four quark systems. We prefer to use the diquark-antidiquark picture to estimate the open charm and bottom tetraquark states in the present work. We find that our results give much higher masses of $sq\bar b\bar q$ tetraquark configuration. It disfavors the assumption of $X(5568)$ as a tetraquark state within the diquark-antidiquark scenario.

This paper is organized as follows.  The relativized quark model is briefly introduced and the masses of diquarks are calculated in Sec.~\ref{diquark}. The masses of tetraquark states and discussions are presented in Sec.~\ref{tetraquark}. Finally, we give a short summary in the last section.

\section{Masses of diquarks}{\label{diquark}}

The Hamiltonian between quark and antiquark in the relativized quark model can be expressed as
\begin{equation}
\tilde{H} = H_0+\tilde{V}(\boldsymbol{p},\boldsymbol{r}), \label{ham}
\end{equation}
\begin{equation}
H_0 = (p^2+m_1^2)^{1/2}+(p^2+m_2^2)^{1/2},
\end{equation}
\begin{equation}
\tilde{V}(\boldsymbol{p},\boldsymbol{r}) = \tilde{H}^{conf}_{12}+\tilde{H}^{cont}_{12}+\tilde{H}^{ten}_{12}+\tilde{H}^{so}_{12},
\end{equation}
where the $\tilde{H}^{conf}_{12}$ includes the spin-independent linear confinement and Coulomb-like interaction, the $\tilde{H}^{cont}_{12}$ is the color contact term, the $\tilde{H}^{ten}_{12}$ is the color tensor interaction, and $\tilde{H}^{so}_{12}$ is the spin-orbit term. $\tilde{H}$ denotes an operator that has taken account of the relativistic effects according to the relativized scheme. The explicit forms of these interactions and the details of this relativization procedure can be found in Ref.~\cite{Godfrey:1985xj}. For the quark-quark interaction in a diquark, the relation $\tilde{V}_{qq}(\boldsymbol{p},\boldsymbol{r})=\tilde{V}_{q\bar q}(\boldsymbol{p},\boldsymbol{r})/2$ is employed since we only consider the $\bar 3$ type diquark in color space. All the model parameters used in our calculations are taken from Ref.~\cite{Godfrey:1985xj}. It should be emphasized that these parameters can describe the low lying meson and baryon spectra well even comparing with the recently experimental data. While for the higher excited states, since the discussions of their assignments do not agree with each other theoretically, and they can be hardly employed for fitting the new parameters. Since only the ground states of the tetraquarks are calculated in the present work, where the parameters about the spin-orbit and tensor forces are not employed, we believe that it is enough to describe the ground tetraquark states with the original parameters.

For the diquark, the $qq$ locates in $S$ wave. The spin-parities of the diquark are $J^P=0^+$ and $J^P=1^+$, named as the scalar diquark and axial diquark, respectively. We use the Gaussian expansion method to solve the Hamiltonian~(\ref{ham}) with $\tilde{V}_{qq}(\boldsymbol{p},\boldsymbol{r})$ potential~\cite{Hiyama:2003cu}. The obtained masses of these diquarks are listed in Table.~\ref{tab1}.

\begin{table}[!htbp]
\begin{center}
\caption{ \label{tab1} The masses of the scalar and axial vector diquarks. $S$ and $A$ denote the scalar and axial vector diquarks, respectively. The brace and bracket correspond to symmetric and antisymmetric quark contents in flavor, respectively.}
\small
\begin{tabular*}{8.5cm}{@{\extracolsep{\fill}}*{3}{p{2.5cm}<{\centering}}}
\hline\hline
 Quark content &  Diquark type & Mass(MeV)\\\hline
 $[u,d]$       &  $S$            & 691 \\
 $\{u,d\}$     &  $A$            & 840 \\
 $[u,s]$       &  $S$            & 886 \\
 $\{u,s\}$     &  $A$            & 992 \\
 $\{s,s\}$     &  $A$            & 1135 \\
 $[c,q]$       &  $S$            & 2099 \\
 $\{c,q\}$     &  $A$            & 2138 \\
 $[c,s]$       &  $S$            & 2230 \\
 $\{c,s\}$     &  $A$            & 2264 \\
 $[b,q]$       &  $S$            & 5451 \\
 $\{b,q\}$     &  $A$            & 5465 \\
 $[b,s]$       &  $S$            & 5572 \\
 $\{b,s\}$     &  $A$            & 5585 \\
 \hline\hline
\end{tabular*}
\end{center}
\end{table}

\section{Masses of tetraquark states}{\label{tetraquark}}

With the diquark listed in Tab.~\ref{tab1}, we can calculate the masses of the tetraquarks regarded as the bound states of the diquark and antidiquark. While the diquark structure is considered, the form factor $F(r)$ emerges in the Coulomb-like one gluon exchange potential, that is, the $-1/r$ term in the quark-antiquark interaction becomes $-F_1(r)F_2(r)/r$  in the diquark-antidiquark interaction. The $F(r)$, which stands for the internal structure of diquark, can be approximated, with a high accuracy, by the following expression~\cite{Ebert:2010af}
\begin{equation}
F(r) = 1-e^{-\xi r -\zeta r^2},
\end{equation}
where the $\xi$ and $\zeta$ are the nonnegative real numbers. It can be found that the form factor satisfies $0\leq F(r)\leq 1$. In our present work, we calculate the tetraquark mass with $F(r)=1$, and we will discuss the effects induced by this form factor qualitatively later on.

\begin{figure}[!htbp]
\includegraphics[scale=0.6]{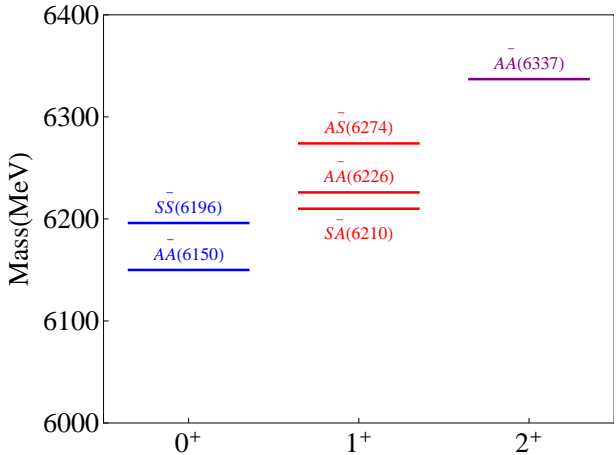}
\vspace{0.0cm} \caption{The predicted mass spectrum of the $sq\bar b \bar q$ tetraquarks.}
\label{mass}
\end{figure}

In the $F(r)=1$ case, the screen effects due to the finite size of the diquark are totally neglected, namely, the diquark is treated as a point-like antiquark or the distance between the diquark and antidiquark is large enough~\cite{Maiani:2004vq,Santopinto:2004hw,Ferretti:2011zz,Santopinto:2014opa,Maiani:2014aja,Maiani:2015vwa,Brodsky:2014xia,Lebed:2015tna,Lu:2016cwr}. The predicted masses of the open charm and bottom tetraquark states in $1S$ ground states are shown in Tab~\ref{tab2}. For the $sq\bar b \bar q$ quark content, we also plot the mass spectrum in Fig.~\ref{mass}. We find that for a certain quark flavor, the lowest state is the $J^P=0^+$ $A\bar A$ type diquark-antidiquark configuration. This order of mass spectrum is different from the results of Ref.~\cite{Ebert:2010af}, in which the $J^P=0^+$ $S\bar S$ type is the lowest one. If the spin-spin interaction is treated perturbatively, it can provide the fine splitting with coefficients of -2, -1, and 2 for the $0^+$, $1^+$, and $2^+$ $A\bar A$ states, respectively, and no fine structure exists for the $S\bar S$ tetraquark. Although the mass of $A$ type diquark is higher than that of $S$ type, the larger fine splitting induced by the spin-spin interaction can reduce $J^P=0^+$ $A\bar A$ state to be the lowest one. The lowest mass of $J^P=0^+$ $sq\bar b \bar q$ state, in our calculation, is 6150 MeV, which is much larger than the mass of $X(5568)$ state. The lowest mass of $J^P=1^+$ $sq\bar b \bar q$ state is 6210 MeV, which is also larger than $m(X(5568)) + m(B^*_s)- m(B_s)$. We see that for both the $0^+$ and $1^+$ cases, the predicted masses are much larger than the experimental data. Hence, our calculated results disfavor the possibility of $X(5568)$ as a tetraquark state within the diquark-antidiquark scenario.

\begin{table}[!htbp]
\begin{center}
\caption{ \label{tab2} The masses of tetraquark states with diquark-antidiquark in ground $1S$ state. A dash denotes this state do not exist.}
\small
\begin{tabular*}{8.5cm}{@{\extracolsep{\fill}}*{4}{p{1.7cm}<{\centering}}}
\hline\hline
 $J^P$ &  Diquark content & Open charm (MeV) & Open bottom (MeV)\\\hline
               &               & $\bm{cq\bar q \bar q}$      & $\bm{qq\bar b \bar q}$    \\
 $0^+$         &  $S\bar S$      & 2729                 & 6063               \\
 $1^+$         &  $S\bar A$      & 2838                 & 6077               \\
 $1^+$         &  $A\bar S$     & 2767                 & 6164               \\
 $0^+$         &  $A\bar A$      & 2575                 & 6046               \\
 $1^+$         &  $A\bar A$      & 2747                 & 6118               \\
 $2^+$         &  $A\bar A$      & 2969                 & 6226               \\\hline
               &               & $\bm{cq\bar s \bar q}$      & $\bm{sq\bar b \bar q}$    \\
 $0^+$         &  $S\bar S$      & 2873                 & 6196               \\
 $1^+$         &  $S\bar A$      & 2957                 & 6210               \\
 $1^+$         &  $A\bar S$      & 2911                 & 6274               \\
 $0^+$         &  $A\bar A$      & 2692                 & 6150               \\
 $1^+$         &  $A\bar A$      & 2866                 & 6226               \\
 $2^+$         &  $A\bar A$      & 3087                 & 6337               \\\hline
               &               & $\bm{cs\bar s \bar q}$      & $\bm{sq\bar b \bar s}$    \\
 $0^+$         &  $S\bar S$      & 3001                 & 6317               \\
 $1^+$         &  $S\bar A$      & 3085                 & 6330               \\
 $1^+$         &  $A\bar S$      & 3035                 & 6394               \\
 $0^+$         &  $A\bar A$     & 2827                 & 6272               \\
 $1^+$         &  $A\bar A$      & 2994                 & 6347               \\
 $2^+$         &  $A\bar A$      & 3207                 & 6456               \\\hline
               &               & $\bm{cs\bar s \bar s}$      & $\bm{ss\bar b \bar s}$    \\
 $1^+$         &  $S\bar A$      & 3201                 & $-$                \\
 $1^+$         &  $A\bar S$      & $-$                  & 6504               \\
 $0^+$         &  $A\bar A$      & 2942                 & 6376               \\
 $1^+$         &  $A\bar A$      & 3111                 & 6455               \\
 $2^+$         &  $A\bar A$      & 3322                 & 6566               \\
\hline\hline
\end{tabular*}
\end{center}
\end{table}

Different from our results, some works claim that they can describe the mass of $X(5568)$ in tetraquark picture. In those calculations, the selected masses of diquarks are lower than ours, and they are not consistently obtained by solving the quark model potentials. For example, a set of parameters with $m_{bq}=5.249~\rm{GeV}$ and $m_{sq}=0.590~\rm{GeV}$ are used to calculate the masses of $sq\bar b q$ tetraquarks, in which the lowest $J^P=0^+$ states is about 150 MeV higher than the $X(5568)$~\cite{Wang:2016tsi,Maiani:2004vq}. If we take those lighter diquark values as the masses of the $A$ type diquarks and solve the relativized Schr\"{o}dinger-type equation, we can also obtain a rather low tetraquark mass of 5672 MeV. However,
our present consistent calculations both for the diquarks and tetraquarks in the realistic potential does not support the $X(5568)$ as a tetraquark state.

Moreover, we know that the masses of $bsu$ flavor baryons $\Xi_b$ and $\Xi_b^*$ are 5794 and 5945 MeV, respectively. It is natural to believe that the $sq\bar b \bar q$ tetraquarks, containing an additional valence quark, should be above or at least around these two baryon states~\cite{Burns:2016gvy}. Another argument is that the tetraquark masses can be estimated by the spin-averaged mass of mesons roughly. This situation occurs in the study of the $XYZ$ masses, where the tetraquark and molecular pictures can both give similar results in the most cases~\cite{Chen:2016qju}. The spin-averaged mass $(3M_V+M_P)/4$ of $B^*$, $B$, $K^*$, and $K$ is 6107 MeV~\cite{Burns:2016gvy}, which is well consistent with the $6150~\rm{MeV}$ ($A\bar{A}$
case) in our present calculation.

When the finite size of diquark is considered, the one gluon exchange interaction between the diquark and antidiquark becomes weaker as well as the spin-spin interaction. The masses of the tetraquarks will increase, while the fine splitting becomes smaller. This situation is the same as the case adopted by Ebert, Faustov, and Galkin, where the mass of $J^P=0^+$ $S\bar S$ type tetraquark is the lowest and the fine splitting is small~\cite{Ebert:2010af}. In the $F(r)=0$ limit, only the linear confining interaction remains, and the three $A\bar A$ type tetraquark states degenerate. Of course, the $X(5568)$ cannot be described as a tetraquark state even the finite size and form factor of the diquark are taken into account.

In present work, we predict many open charm and bottom tetraquark states within the diquark-antidiquark scenario by solving the Schr\"{o}dinger-type equation. It should be noted that only the mass spectra cannot ensure the existences of these states, and their production mechanisms and decay behaviors should also be investigated simultaneously. The strong decay behaviors are more essential, since the much broader structures cannot form or be detected. In fact, we predict the lowest $sq\bar b\bar q$ state is 6150 MeV, which is much higher than the $B_s \pi$ and $B \bar K$ thresholds. Due to the large phase space, the predicted tetraquark states may fall apart immediately. Further studies on these tetraquark states are needed both theoretically and experimentally.

\section{Summary}{\label{Summary}}

In this work, we study the masses of open charm and bottom tetraquark states in the diquark-antidiquark pictures using the relativized quark model proposed by Godfrey and Isgur. The diquark and antidiquark masses are obtained with the relativized potential, which is the half of $q\bar q$ interaction. Then, the diquark and antidiquark are regarded as the usual antiquark and quark, respectively. The form factor, simulating the diquark (antiquark) internal structure, is neglected in our calculations. This
assumption means the diquark  is treated as point-like or the distance between diquark and antidiquark is large enough.

The masses of the tetraquark states are obtained by solving the Schr\"{o}dinger-type equation between diquark and antidiquark. We find the masses of $sq\bar b\bar q$ tetraquark configuration are much higher than that of $X(5568)$, which disfavors the possibility of $X(5568)$ as a tetraquark state within the diquark-antidiquark scenario. The effects induced by the
form factor and the finite size of the diquark are qualitatively analyzed. We expect that further experimental information can reveal the nature of the signal observed by D0 collaboration.

\bigskip
\noindent
\begin{center}
{\bf ACKNOWLEDGEMENTS}\\
\end{center}

We would like to thank Fei Huang for valuable discussions, and Tim Gershon for reminding us the LHCb results and giving helpful interpretation. This project is supported by the National Natural Science Foundation of China under Grants No.~10975146, and No.~11475192. The fund provided by the Sino-German CRC 110 ``Symmetries and the Emergence of Structure in QCD" project (NSFC Grant No. 11261130311) is also appreciated.
YBD thanks Alexander von Humboldt Foundation and the Institute of Theoretical Physics, University of Tuebingen for
the warm hospitality.

\end{document}